\documentclass[%
 reprint,
 amsmath,amssymb,
 aps,showpacs,superscriptaddress
]{revtex4-1}

\usepackage{graphicx}
\usepackage{dcolumn}
\usepackage{bm}
\usepackage{color}

\newcommand{\beqar}{\begin{eqnarray}}
\newcommand{\eeqar}{\end{eqnarray}}

\newcommand{\bcen}{\begin{center}}
\newcommand{\ecen}{\end{center}}

\newcommand{\eps}{\varepsilon}
\newcommand{\lam}{\lambda}
\newcommand{\bra}[1]{\left< #1 \right|}
\newcommand{\ket}[1]{\left| #1 \right>}

\newcommand{\f}[2]{\frac{#1}{#2}}
\renewcommand{\b}[1]{\left({#1}\right)}

\renewcommand{\sb}[1]{\left[{#1}\right]}
\newcommand{\mean}[1]{\langle {#1} \rangle}

\newcommand{\ra}{\rightarrow}

\newcommand{\braket}[2]{ \left< #1 \right| \left| #2 \right>}

\begin{document}

\preprint{APS/123-QED}
\title{Open system dynamics from thermodynamic compatibility }

\author{Roie Dann}
\email{roie.dann@mail.huji.ac.il}
\affiliation{The Institute of Chemistry, The Hebrew University of Jerusalem, Jerusalem 9190401, Israel}%
\author{Ronnie Kosloff}%
\email{kosloff1948@gmail.com}
\affiliation{The Institute of Chemistry, The Hebrew University of Jerusalem, Jerusalem 9190401, Israel}%

\date{\today}
\begin{abstract}
Thermodynamics entails a set of mathematical conditions on quantum Markovian dynamics. In particular, strict energy conservation between the system and environment implies that
the dissipative dynamical map commutes with the unitary system propagator.
Employing spectral analysis we prove the general form of the ensuing master equation. We compare this result to  master equations obtained from standard microscopic derivations.  
The obtained formal structure can be employed 
to test the compatibility of approximate
derivations with thermodynamics.
For example, it designates that global master equations are the compatible choice. The axiomatic approach sheds light on the validity of the secular approximation in microscopic derivations, the form of the steady state in heat transport phenomena, and indicates the lack of exceptional points in the dynamics of open quantum systems. 
\end{abstract}

\maketitle

\section{Introduction}
\label{sec:intro}
The generic dynamics of an open quantum system directs the system towards a thermal equilibrium state.
Theoretically, this phenomena is described within a physical theory which constitutes a dialog between quantum mechanics and thermodynamics. As most physical theories, it is formulated in terms of a set of postulates.  Our present goal is to identify a minimal set of postulates, based on both thermodynamics and quantum mechanics principles, and infer the system dynamics from the axiomatic description. 

Open quantum system dynamics are naturally described within von Neumann’s formulation of quantum mechanics \cite{von2018mathematical}. In this framework the composite system
state, including the primary system and the environment, is described by the composite density operator $\hat{\rho}$.
The underlying assumption of this framework is that the composite system is prepared in an uncorrelated state. This restriction is motivated
by the thermodynamic properties of quantum systems,
as it enforces a positive quantum entropy production \cite{landi2020irreversible,esposito2010entropy}. 

Thermodynamic principles impose a number of additional restrictions.
Conservation of energy, or the $I$-law implies that when no external force is applied, a time-independent Hamiltonian 
\begin{equation}
 \hat{H}=\hat{H}_S+\hat{H}_{SE}+\hat{H}_E~~,
\label{eq:Hamiltonian}
\end{equation}
generates the composite unitary dynamics. The  Hamiltonian is composed of the system and environment (rest of the universe) free Hamiltonians $\hat{H}_S$ and $\hat{H}_E$, and an interaction term $\hat{H}_{SE}$.

Focusing on the system, the reduced dynamics is obtained by tracing out the environment, leading to a completely positive trace preserving map (CPTP)
\begin{multline}
\hat{\rho}_S\b t={\Lambda}_t\sb{\hat{\rho}_S\b 0}\\=\text{tr}_E\b{\hat{U}\b{t,0}\hat{\rho}_S\b 0\otimes \hat{\rho}_E\b 0\hat{U}^\dagger\b{t,0}}~,   
\label{eq:cal_Lambda_t}
\end{multline}
where $\text{tr}_E$ signifies the partial trace over the environment degrees of freedom and $\hat{U}$ is the evolution operator of the composite system.

Another thermodynamic idealization states that the environment is large, stationary and characterized by a temperature. Such behaviour can only be obtained in the quantum setting within the Markovian regime.  If a separation of timescales between the environment and system exists, the memory in the environment decays rapidly, and the environment can effectively remain in a stationary state. Mathematically, this means that the map satisfies the semi-group property  $\Lambda_t=\Lambda_{t-s}\Lambda_s$, which in turn allows expressing the map in terms of the dynamical semi-group generator $\Lambda_t=e^{{\cal L} t}$
\cite{davies1976quantum,alicki2007quantum}. We will address the most general form of this generator. This form was originally described by Gorini, Kossakowski as well as Sudarshan \cite{gorini1976completely}, and separately by Lindblad \cite{lindblad1976generators}, and termed the GKLS form.

The $II$-law of thermodynamics implies that the system monotonically approaches
the steady state. In the quantum setting this is manifested by the contracting property of the map for any two states
\begin{equation}
    {\cal S}\b{\hat{\rho}_S||\hat{\rho}_S'}\geq{\cal S}\b{\Lambda_t\hat{\rho}_S||\Lambda_t\hat{\rho}_S'}
    \label{eq:divergence}
\end{equation}
where $\cal S$ is the relative entropy (or Kullback–Leibler divergence), which is a measure of the distance between two states \cite{lindblad1974expectations}.
In addition, the $0$'th law of thermodynamics suggests that the map has a single fixed point $\hat{\rho}_S^{eq}$. Such a state is an invariant of the map and constitutes an eigenoperator with eigenvalue $\bf 1$. Invariance of $\hat{\rho}_S^{eq}$ and Eq. \eqref{eq:divergence} then imply that successive applications of the map will monotonically lead to equilibrium \cite{frigerio1977quantum}
\begin{equation}
    {\cal S}\b{\hat{\rho}_S||\hat{\rho}_S^{eq}} \geq     {\cal S}\b{\Lambda_t\hat{\rho}_S||\hat{\rho}_S^{eq}}~~.
    \label{eq:contracting_map}
\end{equation}
This relation is more familiar in its differential form, known as Spohn's inequality \cite{spohn1978irreversible} \footnote{Equation \eqref{eq:contracting_map} interpreted as the quantum version of the $II$-law \cite{alicki1979quantum}.}.
The fixed point of the map together with the CPTP property impose strict conditions on the system dynamics. 


A final postulate emerges from the thermodynamic principle which assumes that energy is not trapped in the interface between the system and environment. Hence, any change of energy in the system is mirrored by a reverse change in the environment. This condition can be formulated by the relation
\begin{equation}
\sb{\hat{H}_{SB},\hat{H}_S+\hat{H}_E}=0~~,    
\label{eq:strict_energy_conservation}
\end{equation}
commonly termed {\emph{strict energy conservation}} \cite{janzing2000thermodynamic}.
This idealization is compatible with the thermodynamic limit, in which the interface energy is negligible with respect to bulk energy.

We will prove that these thermodynamic constraints impose a strict condition on the dynamical map: the open-system dynamical map $\Lambda$ commutes with the unitary dynamical map ${\cal U}_S$, associated with the isolated system evolution 
\begin{equation}
    {\cal{U}}_S\sb{\Lambda\sb{\hat{\rho}_S}}={\Lambda\sb{{\cal{U}}_S\sb{\hat{\rho}_S}}}~~.
    \label{eq:maps_commute}
\end{equation}
Here ${\cal U}_S=e^{-i{\cal{H}}_S t}$, where ${\cal H}_S$ is the generator of the unitary map ${\cal H}_S\sb{\hat{\rho}_S\b t}={\hbar^{-1}}\sb{\hat{H}_S,\hat{\rho}_S\b t}$. We coin this relation as the ``commutativity property of the maps" and the related theorem by ``Theorem 1".
The property that the two dynamical maps commute, Eq. \eqref{eq:maps_commute}, suggests that their generators also commute
\begin{equation}
    {\cal{H}}_S\sb{{\cal L}\sb{\hat{\rho}_S}}={\cal L}\sb{{\cal{H}}_S\sb{\hat{\rho}_S}}~~.
\end{equation}
In turn, 
since ${\cal{H}}_S$ and $\cal{L}$ commute, they share a common basis of eigenoperators. 

The question arises: How do the spectral properties, imposed by thermodynamics, restrict the structure of the GKLS master equation?
This paper is dedicated to the analysis of this issue, i.e., the dynamical implications  associated with thermodynamic compatibility.
We find that these conditions are sufficient to limit the structure of the master equation to a very restrictive form, without introducing any further approximations. 

\section{Framework and representation}
\label{sec:framework_and_representation}
The exponential form of map $\Lambda_t$ in the Markovian regime leads to the differential form of the master equation
\begin{equation}
    \f{d}{dt}\hat{\rho}_S\b t ={\cal L}\sb{ \hat{\rho}_S\b t}~~.
    \label{eq:dynamics}
\end{equation}
Generally, $\cal L$ is composed of unitary and dissipative terms
\begin{equation}
{\cal{L}}=-i{\cal H}+{\cal{D}}~~.
\label{eq:L_decompos}
\end{equation}
This decomposition is not unique \cite{breuer2002theory}, nevertheless, the present analysis relies on the commutation  properties of the two terms which are independent of the decomposition \eqref{eq:commutivity}.
The dissipative term $\cal D$ of the GKLS formulation has the structure
\begin{equation}
    {\cal{D}}\sb{\bullet} = \sum_k \gamma_k\b{ \hat{L}_k\bullet \hat{L}_k^\dagger-\f{1}{2}\left\{\hat{L}_k^\dagger \hat{L}_k,\bullet \right\}}~~,
    \label{eq:gen_dissipator}
\end{equation}
where the kinetic coefficients $\gamma_k$ are non-negative and $\hat{L}_k$ are called the Lindblad jump operators. In this structure, $\gamma_k$ and $\hat{L}_k$ remain unspecified and in general depend on the details of the system and environment. 

For a specific physical scenario, $\cal{L}$ 
is typically obtained  from  a complete quantum description, starting from  Eq. \eqref{eq:Hamiltonian}.  A microscopic derivation is preformed, leading to a master equation of the GKLS form \cite{davies1974markovian,davies1976quantum,breuer2002theory}. Therefore, guaranteeing a CPTP map and consistency with the postulates of quantum mechanics.  This procedure requires a number of approximations, which determine a restricted regime of application.
There is a long history of microscopic derivations \cite{nakajima1958quantum,zwanzig1960ensemble,davies1974markovian,alicki1977markov,diosi1993calderia,lidar2001completely,daffer2004depolarizing,shabani2005completely,maniscalco2006non,breuer2008quantum,whitney2008staying,piilo2008non,alicki2012periodically,szczygielski2013markovian,majenz2013coarse,albash2012quantum,muller2017deriving,smirnov2018theory,dann2018time,mccauley2019completely,mozgunov2020completely,nathan2020universal}, where different derivations may differ by  their regime of validity, the kinetic coefficients and the Lindblad operators.

In the following study we approach this problem by a different methodology. We introduce three additional conditions (postulates), motivated by thermodynamic principles,  and show that these determine the master equation up to a scaling of independent kinetic coefficients. 

In the derivations we utilize a representation of open system dynamics known as Liouville or Hilbert-Schmidt space \cite{fano1957description,alicki1976detailed,gohberg1990hilbert,sarandy2005adiabatic,sarandy2006abelian,albert2016geometry,shabani2005completely}. It is defined as the space of all system operators $\{\hat{X}\}$ endowed with an inner product $\b{\hat{X}_i,\hat{X}_j}\equiv\text{tr}\b{\hat{X}_i^\dagger\hat{X}_j}$. If the dimension of the system's Hilbert space is $N$, then the Liouville space is of dimension $N^2$ \footnote{Even when the dimension of the system's Hilbert space is infinite, the system dynamics can still be represented in Liouville space if the dimension of the operator algebra is finite. For example, the dynamics of a Gaussian state of the quantum harmonic oscillator can be described by a closed operator of algebra, and consequently as a time-dependent finite vector in Liouville space \cite{rezek2006irreversible}. }. In this space the density operator is represented by an $N^2\times1$ vector which defines the mapping between the two spaces.   Superoperators such as $\Lambda$ and $\cal D$, which constitute linear transformations of system operators, are represented by $N^2\times N^2$ matrices.  We denote superoperators in Liouville space by a wide-tilde superscript ($\widetilde{{\Lambda}}$ and $\widetilde{{\cal D}}$). The matrix vector formulation of the open system dynamics provides a simple framework to analyze open system dynamics.

\section{Thermodynamically motivated postulates}
\label{sec:preliminary_conditions}

In the present framework, the fundamental postulates of quantum mechanics are supplemented with four thermodynamically motivated postulates: 
\begin{enumerate}
    \item The dynamical map $\Lambda_t$ is Markovian, satisfying the semi-group property.
    \item The environment remains in a stationary state with respect to the environment's free Hamiltonian $\hat{H}_E$. 
    \item The fixed point of the dynamical map is a thermal state $\hat{\rho}_S^{eq}=\hat{\rho}_S^{th}=Z_{S}^{-1} e^{-\beta\hat{H}_S}$, where $Z_S$ is the partition function and $\beta$ is the inverse temperature of the environment.
    \item  The composite system satisfies {\emph {strict energy conservation}} between the system and environment: $\sb{\hat{H}_{SE},\hat{H}_S+\hat{H}_E}=0$.
\end{enumerate}
These postulates impose limitations on the dynamics of an open quantum system.
An immediately consequence of postulate 4 is that any state which is of the form $f\b{\hat{H}_S+\hat{H}_E}$ is an invariant of the global dynamics, where $f$ is an arbitrary analytic function  \cite{janzing2000thermodynamic,brandao2013resource,lostaglio2019introductory}. When the environment remains approximately in a thermal state $\hat{\rho}_E^{th}=Z_{E}^{-1}e^{-\beta\hat{H}_E}$ throughout the dynamics (a specific case of postulate 2), the global invariant must be $\hat{\rho}^{th}=Z^{-1}e^{-\beta\b{\hat{H}_S+\hat{H}_E}}$, which is consistent with the $0$'th law. In turn, the associated reduced system state is 
$\text{tr}_E\b{\hat{\rho}^{th}}=\hat{\rho}_S^{th}$, which motivates the {\emph{a priori}} postulation of postulate 3.



\section{Restricted structure of the master equation}
\label{sec:restricted_sturcture_of_ME}

We now show that these four postulates restrict the form of the generator of the quantum dynamical semi-group.  They set the Lindblad operators $\hat{L}_k$ as  the eigenoperators of the free propagator, as well as the ratio between pairs of kinetic coefficients.


We start the analysis by studying the commutativity properties of dynamical maps, and their generators.

\subsection{Commutation properties of dynamical maps and generators }
\label{subsec:commutivity}
The dynamical map associated with the composite dynamics and the map of the isolated system satisfy the following theorem.
\paragraph*{Theorem 1} 
Let $\hat{H}$ (Eq. \eqref{eq:Hamiltonian}) be the time-independent Hamiltonian of the composite system,
with $\sb{\hat{H}_{SE},\hat{H}_{S}+\hat{H}_{E}}=0$, and let the initial state $\hat{\rho}_E\b 0$ be a stationary state of  $\hat{H}_E$
 then  the dynamical maps $\Lambda_t$, Eq. \eqref{eq:cal_Lambda_t} and  ${\cal U}_{S}\sb{\hat{\rho}_{S}\b 0}=\hat{U}_{S}\b{t,0}\hat{\rho}_{S}\b 0\hat{U}_{S}^{\dagger}\b{t,0}$ commute, where $\hat{U}_S\b{t,0}=e^{-i \hat{H}_S t/\hbar}$ is the free propagator of the system and $\hat{U}\b{t,0}=e^{-i\hat{H}t/\hbar}$.
\paragraph*{Proof} 
We first introduce some notations: The free propagators of the environment and composite (uncoupled) system are given by $\hat{U}_E\b{t,0}=e^{-i\hat{H}_E t/\hbar}$ and 
$\hat{U}_{SE}\b{t,0}=e^{-i\b{\hat{H}_S+\hat{H}_E} t/\hbar}$, moreover, the spectral decomposition of the environment Hamiltonian reads $\hat{H}_E =\sum_i c_i \ket{\chi_i} \bra{\chi_i}$.
Since the initial state of the environment is stationary with respect to $\hat{H}_E$, it can also be expressed as $\hat{\rho}_E\b 0=\sum_{i}\lam_{i}\ket{\chi_{i}}\bra{\chi_{i}}$. To simplify the notation, in this proof we emit the time-dependence of the propagators and maps, nevertheless, it should be clear that they induce a time translation from initial time ($t'=0$) to time $t'=t$.

Utilizing the spectral decomposition of the environment's initial state any quantum dynamical map can be expressed in a Kraus form \cite{kraus1971general}
\begin{equation}
    \hat \rho_{S}\b t=\sum_{ij}\hat{K}_{ij}\hat \rho_{S}\b 0 \hat K_{ij}^{\dagger}~~, 
    \label{eq:Kraus_form}
\end{equation}
where $\hat{K}_{ij}=\sqrt{\lam_{i}}\bra{\chi_{j}}\hat U\b{t,0}\ket{\chi_{i}}$ with $\sum_{ij}\hat{K}_{ij}^\dagger \hat{K}_{ij}=\hat{I}_S$. In the Heisenberg representation the dynamical map becomes $\hat{O}^H_S\b t ={\Lambda}^*\sb{\hat{O}_S} = \sum_{ij}\hat{K}_{ij}^{\dagger}\hat O_{S}\b 0 \hat K_{ij} $, where the superscript $H$ and asterisk designate operators and superoperators (dynamical maps) in the Heisenberg representation and $\hat{O}_S$ is a system operator. 

Using the Kraus representation the product of dynamical maps is explicitly expressed as
\begin{multline}
    {\cal U}_{S}^*\sb{{{\Lambda}}^{*}\sb{\hat O_{S}}}=\hat U_{S}^{\dagger}\b{\sum_{ij}\hat K_{ij}^{\dagger}\hat O_{S}\hat K_{ij}}\hat U_{S}\\=\sum_{i}\lam_{i}\bra{\chi_{i}}\hat U_{S}^{\dagger}\hat U^{\dagger}\hat O_{S}\sum_{j}\ket{\chi_{j}}\bra{\chi_{j}}\hat U \hat U_{S}\ket{\chi_{i}}\\=\sum_{i}\lam_{i}\bra{\chi_{i}}\hat U_{S}^{\dagger}\hat U^{\dagger}\hat O_{S}\hat U \hat U_{S}\ket{\chi_{i}}~~,
\end{multline}
where the second equality is achieved by identifying the environment identity operator $\hat{I}_E = \sum_{j}\ket{\chi_{j}}\bra{\chi_{j}}$. Inserting the identity operator $\hat{U}_E\hat U_E^\dagger = \hat{I}_E $ twice we obtain 
\begin{multline}
     {\cal U}_{S}^*\sb{ {\Lambda}^{*}\sb{\hat O_{S}}}=\sum_{i}\lam_{i}\bra{\chi_{i}}\hat U_{E}\hat{U}_{SE}^{\dagger}\hat U^{\dagger}\hat O_{S}\hat U\hat U_{SE}\hat U_{E}^{\dagger}\ket{\chi_{i}}\\
     =\sum_{i}\lam_{i}\bra{\chi_{i}}\hat{U}_{SE}^{\dagger}\hat U^{\dagger}\hat O_{S}\hat U\hat U_{SE}\ket{\chi_{i}}~~.
\end{multline}
 The second equality is obtained by utilizing the eigenvalue equation  $\hat{U}_E\ket{\chi_i} = e^{-i c_i t/\hbar}\ket{\chi_i}$ for the eigenstates $\{\ket{\chi_i}\}$. Next, strict energy conservation implies that $\sb{\hat{H},\hat{H}_S+\hat{H}_E}=0$, which in turn suggests that the associated propagators satisfy $\sb{\hat{U},\hat{U}_{SE}}=0$. This relation leads to the final form 
 \begin{equation}
     {\cal U}_{S}^*\sb{{ {\Lambda}^{*}}\sb{\hat O_{S}}}  =\sum_{i}\lam_{i}\bra{\chi_{i}}\hat U^{\dagger}\hat{U}_{SE}^{\dagger}\hat O_{S}\hat U_{SE}\hat U\ket{\chi_{i}}~~.
     \label{eq:direct_product}
 \end{equation}
Following a similar derivation the product in reverse order of the dynamical maps gives
\begin{multline}
    {\Lambda}^{*}\sb{{{\cal U}}_{S}^*\sb{\hat O_{S}}}=\sum_{ij}\hat K_{ij}^{\dagger}\hat U_{S}^{\dagger}\hat O_{S}\hat U_{S}\hat K_{ij}\\ =\sum_{i}\lam_{i}\bra{\chi_{i}}\hat U^{\dagger}\hat U_{SE}^{\dagger}\hat U_{E}\hat O_{S}\hat U_{E}^{\dagger}\hat U_{SE}U\ket{\chi_{i}}\\ =\sum_{i}\lam_{i}\bra{\chi_{i}}\hat U^{\dagger}\hat U_{SE}^{\dagger}\hat O_{S}\hat U_{SE}U\ket{\chi_{i}}~~,
    \label{eq:reverse_product}
\end{multline}
where the last equality stems from the commutativity of local operators of the  system and environment $\sb{\hat{U}_E,\hat{O}_S} = 0$. 

Finally, Eqs. \eqref{eq:direct_product} and \eqref{eq:reverse_product} imply the desired result 
\begin{equation}
    {\cal U}_{S}^*\sb{{{\Lambda}^{*}}\sb{\hat O_{S}}}={\Lambda}^{*}\sb{{\cal {\cal U}}_{S}^*\sb{\hat O_{S}}}~~_\blacksquare
\end{equation}
From the equivalence of the Schr\"odinger and Heisenberg representations we can infer that $\Lambda$ and ${\cal{U}}_S$ commute.  In Liouville space this relation is simply expressed as $\sb{\widetilde{\cal{U}}_S,\widetilde{\Lambda}} = 0$.

This result is quite general, as the proof employed postulate 4 of strict energy conservation and the property that environment is stationary. It did not assume the maps satisfy the semi-group property. Thus, the property of commuting maps holds for arbitrary coupling and a wide range of environments of any size (finite or infinite). Moreover, the dynamics may be  non-Markovian \cite{megier2020interplay}. 

\subsubsection*{Consequences of the commutation relation $\sb{\widetilde{\cal{U}}_S,\widetilde{\Lambda}} = 0$ on the semi-group generators }
If  $\Lambda$ is semi-group map the generator is defined by   ${\cal L}\sb{\hat{\rho}_S} =\lim_{\eps\ra0}\f{1}{\eps}\b{{\Lambda}_\eps\sb{\hat{\rho}_S}-\hat{\rho}_S}$ \cite{davies1976quantum}. This definition together with Theorem 1 implies the commutativity of  ${\cal U}_S$ with the generator of the dynamical map $\cal L$.
We define the Lamb-shift as ${\cal H}-{\cal H}_S$, which typically commutes with ${\cal H}_S$ and $\cal D$ \cite{breuer2002theory}. Alternatively, the lamb-shift can be incorporated into $\hat{H}_S$, which infers that ${\cal H}={\cal H}_S$. The ambiguity in the system energy results from the possibility to account a part of the interface energy within the system Hamiltonian.
This property and the fact that  ${\cal{H}}_S$ is an analytical function of ${\cal{U}}_S$ (time-independent Hamiltonian), implies that ${\cal{H}}_S$  commutes with $\cal D$ and $\cal L$, or alternatively
\begin{equation}
    \sb{\widetilde{\cal H}_S,\widetilde{\cal D}}=0~~~~,~~~~\sb{\widetilde{\cal H}_S,\widetilde{\cal L}}=0~~.
\label{eq:commutivity}
\end{equation}

In general $\widetilde{\cal L}$ is non-Hermitian which does not guarantee a spectral decomposition \cite{sarandy2005adiabatic,sarandy2006abelian}. However,
these relation imply that $\widetilde{\cal L}$ and $\widetilde{\cal D}$ are diagonalizable (normal).


\subsection{Identification of the Lindblad operators and relations between kinetic coefficients}

Building upon the implications of theorem 1, we are now ready to determine the Lindblad jump operators and show that the initial conditions imply detailed balance between a pair of dependent coefficients.

Commutativity of superoperators indicates that they share a set of eigenvectors in Liouville space and associated eigenoperators in Hilbert space. 
It is convenient to classify the eigenoperators of ${\cal U}_S$ into two classes: {\emph{invariant}} and {\emph{non-invariant operators}}. The {\emph{invariant}} operators satisfy ${\cal{U}}_S\hat{G}_k=\hat{G}_k$ and therefore have degenerate eigenvalues. As a result, they can be spanned by the energy projection operators of $\hat{H}_S$: $\{\hat{\Pi}_n=\ket{n}\bra{n}\}$. The {\emph{non-invariant}} operators are transition operators between energy states $\{\hat{F}_{nm}=\ket{n}\bra{m}\}$, with $n\neq m $. They satisfy an eigenvalue type equation ${\cal{U}}_S\hat{F}_{nm}=e^{-i\omega_{nm}t}\hat{F}_{nm}$, where $\omega_{nm}=\b{\eps_m-\eps_n}/\hbar$ is the Bohr frequency between energy levels $\ket{n}$ and $\ket{m}$. In the following, we replace the double index by a single index $k$ which runs over all the transition operators. 
Note that $\{\hat{G}\}$ are only invariants of the unitary map, to emphasize this property we refer to them as the {\emph{unitary invariant operators}} and $\{\hat{F}\}$ as the {\emph{unitary non-invariant}}.    

To simplify the analysis we first consider the case when there is no degeneracy in the spectrum of ${\cal U}_S$ except for the unitary invariant subspace. This corresponds to an energy spectrum where no two Bohr frequencies are the same.

We base the present analysis on a standard derivation (including no further assumptions), which starts from the dynamical map, Eqs. \eqref{eq:cal_Lambda_t} and \eqref{eq:Kraus_form},  and leads to a master equation of the GKLS form, Eqs. \eqref{eq:dynamics} \eqref{eq:gen_dissipator} \cite{gorini1976completely,breuer2002theory}.  
The construction includes choosing an orthonormal operators basis $\{\hat{S}\}$ for the system's Liouville space, such that one of the basis operators is chosen to be proportional to the identity and all other operators are traceless. The generator of the quantum dynamical semi-group is then expressed in terms of $\{\hat{S}\}$, leading to (Cf. Appendix \ref{apsec:construction_from_map})
\begin{equation}
    {\cal{D}}^*\sb{\bullet} = \sum_{i,j=1}^{N^2-1}a_{ij}\b{  \hat{S}_j^\dagger\bullet\hat{S_i}-\f{1}{2}\{\hat{S}_j^\dagger\hat{S}_i,\bullet \}}
    \label{eq:gen_GKS}
\end{equation}
where the coefficient matrix $a=\sb{a_{ij}}$  is Hermitian and positive.
We choose the basis $\{\hat{S}\}$ to be composed of the transition operators $\{\hat{F}\}$ and orthonormal operators from the unitary invariant subspace, $\{\hat{G}\}$, these can be chosen as matrices which are proportionate to the diagonal Gell-Mann matricies \cite{gell2010symmetries,georgi2018lie} . We denote this basis by $\mathfrak{F}$. In the general form of Eq. \eqref{eq:gen_GKS}, all the possible
combinations of $\hat{S}_i,\hat{S_j}\in \mathfrak{F}$ exist. However, only certain terms are compatible with the relation \eqref{eq:commutivity} and the eigenoperator condition ${\cal{D}}^*\sb{\hat{F}_k}\propto \hat{F}_k$. 

The  unitary invariant operators $\{G\}$ are generally a linear combination of the energy projection operators $\hat{G}_j = \sum_k g_{jk}\Pi_k$. We can therefore express ${\cal{D}}^*$ in the form of Eq. \eqref{eq:gen_GKS} with jump operators from the basis ${\mathfrak{k}}=\{\hat{F}\}\bigcup\{\hat{\Pi}\}$. Substituting the transition operators  $\hat{F}_k$ into $\cal{D}^*$ and demanding that the eigenoperator condition holds leads to following master equation (Cf. Appendix \ref{apsec:restrictions_on_the_master_eq}) 
\begin{multline}
    {\cal{D}}^*\sb{\bullet} = \sum_{k=-M}^{M}\gamma_{k}\b{  \hat{F}_k^\dagger\bullet\hat{F}_k-\f{1}{2}\left\{\hat{F}_k^\dagger\hat{F}_k,\bullet \right\}}\\+
    \sum_{i,j=1}^{N}\alpha_{ij}\b{\hat{\Pi}_i\bullet \hat{\Pi}_j-\f{1}{2}\left\{\hat{\Pi}_i\hat{\Pi}_j,\bullet \right\}}~~,
    \label{eq:cal_D_pre}
\end{multline}
where $M=N\b{N-1}/2$, $\hat{F}_k^\dagger=\hat{F}_{-k}$, $\gamma_k=a_{kk}$ with $\gamma_0=0$.

A final restriction on the structure of $\cal D$ can be imposed by utilizing postulate 3, which postulates that  $\hat{\rho}_S^{th}=Z^{-1}_Se^{-\beta\hat{H}_S}$ is the fixed point of the dynamical map (Sec. \ref{sec:preliminary_conditions}). This condition determines that dependent kinetic coefficients satisfy detailed balance
\begin{equation}
\gamma_{k}/\gamma_{-k}=e^{-\hbar \omega_k\beta}~~,  
\label{eq:detailed_balance}
\end{equation}
where $\omega_k$ is the Bohr frequency of the energy gap associated with transition operator $\hat{F}_k$. We derive this by substituting $\hat{\rho}_S^{th}$ into the general form of $\cal L$ Eqs. \eqref{eq:dynamics} and \eqref{eq:cal_D_pre} and demanding that ${\cal L}\sb{\hat{\rho}_S^{th}}=0$ (Cf. Appendix \ref{apsec:kinetic_coeff}).

The coefficient matrix $\alpha=\sb{\alpha_{ij}}$ is a real symmetric matrix (Cf. Appendix \ref{apsec:properties_of_alpha}), therefore, it is diagonalizable by an orthogonal matrix $Q$: $Q^T \alpha Q=\text{diag}\b{\lam_1,\dots,\lam_N}$. Introducing a new set of Hermitian operators $\hat{V}_n=\sum_{n}Q_{ni}\hat{\Pi}_n$ and expressing  Eq. \eqref{eq:cal_D_pre}  in terms of the operators $\{\hat{V}\}$ leads to the final form
\begin{multline}
    {\cal{D}}^*\sb{\bullet} = \sum_{k=-M}^{M}\gamma_{k}\b{  \hat{F}_k^\dagger\bullet\hat{F}_k-\f{1}{2}\left\{\hat{F}_k^\dagger\hat{F}_k,\bullet \right\}}\\-
    \sum_{n=1}^{N}\lambda_{n}\sb{\hat{V}_n,\sb{\hat{V}_n,\bullet}}~~.
    \label{eq:cal_D_final}
\end{multline}

This result was obtained under the assumption that the Bohr frequencies of $\hat {H}_S$ are non degenerate. When a degeneracy exists, the commutation relations Eq. \eqref{eq:commutivity} does not imply that the eigenoperators associated with a degeneracy $r$, $\{\hat{F}^{r}\}$ are also eigenoperators of $\cal D$. Generally, the eigenoperators of $\cal D$ are a linear combination $\hat{Y}_k^{r}=\sum_i y^r_{ki} \hat{F}_{i}^{r}$ of the degenerate eigenoperators. In this case, the Lindblad jump operators are generally a combination of $\hat{F}_i^{r}$'s. As a consequence, the master equation may include cross terms of the form $\hat{F}_i^{r}\bullet\hat{F}_j^{r}$ with $i\neq j$. 
 
 The dissipator (Eq. \eqref{eq:cal_D_final}) is composed from two contributions. The first sum induces transitions between eigenstates of $\hat{H}_S$, with propbabilities that satisfy detailed balance Eq. \eqref{eq:detailed_balance}. While the second term describes pure-dephasing, which means loss of coherence between energy eigenstates. Utilizing the spectral decomposition of the system Hamiltonian, the dephasing term can be expressed as a sum of  double commutators of functions of the Hamiltonian $-\sum\sb{f\b{\hat{H}_S},\sb{f\b{\hat{H}_S},\bullet}}$.

\section{Critical analysis of the physical idealizations }
\label{sec:critical_analysis}

The time-reversal symmetry breaking is the main conflict between thermodynamics and quantum mechanics. The theory of open quantum systems resolves this conflict by imposing a partition between a small system and a large environment \cite{lindblad1996existence}. 
In the present study we focus on an isothermal partition which allows heat transport, while maintaining the integrity of the system and environment.
This idealization is reflected by postulates 1 and 4 in Sec. \ref{sec:preliminary_conditions}.

The validity of this idealization has to be confronted with reality. Typically, open system processes are Markovian nature. Even for non-Markovian dynamics the common approach is to embed the system in a larger Markovian framework \cite{li2018concepts,ishizaki2005quantum}. This motivates the {\emph {a priori}} postulation of the semi-group property. The postulate contradicts the principle of unitary evolution,   nevertheless, up to a coarse grained timescale it fits the observed reality. In addition, the Markovian property implies that the composite state must remain separable throughout the evolution \cite{lindblad1996existence}, which enables a local description of the system.

Our construction relies on a  CPTP map which leads to a GKLS form. The complete positively property is related to the separability of the composite state. This assumption has been criticized as non-physical \cite{pechukas1994reduced,dominy2016beyond}. 
In respond to the criticism, justifications for the CPTP property have been presented \cite{alicki1995comment,lindblad1974expectations}.

Strict energy conservation condition is also a thermodynamic idealization and may seem at first too restrictive or nonphysical. We will next show that when the dynamics are Markovian, the condition is effectively implicitly included when the unitary and dissipative maps commute.

Consider an open system with dynamics of a Markovian nature which originate from a general composite Hamiltonian $\hat{H}
'$ with an arbitrary interaction. In general, $\sb{\hat{H}',\hat{H}_{S}}=\hat{X}$,  for some global operator $\hat{X}$. The semi-group property then implies that $\Lambda_{n\tau}=\b{\Lambda_{\tau}}^n$, where $n\in \mathbb{N}$. This apparently cannot be correct for an arbitrary {\emph{coarse-graining time}} $\tau$, since for small enough $\tau$ the Markovian assumption breaks. Therefore, in the present description $\tau$ must be greater (on the order of) the timescale associated with the decay of correlations in the environment. Typically, in the Markovian regime the environment's intrinsic timescale is the fastest one: $\tau\ll \hbar/||\hat{H}_{S}||,\hbar/||\hat{H}_{SE}||$. Hence, the total propagator can be approximated by 
\begin{equation}
    \hat{U}_S\b{\tau,0}\approx\hat{I}-\f i{\hbar}\hat{H}_{S}\tau~~~,~~~
     \hat{U}\b{\tau,0}\approx\hat{I}-\f i{\hbar}\hat{H}'\tau~~.
     \label{eq:infit_propagators}
\end{equation}
Utilizing these relations, the commutation relation of the ``infinitesimal" maps becomes
\begin{multline}
   \Lambda_\tau\sb{{\cal U}_S\b{\tau,0}\sb{\hat{\rho}}}-{\cal U}_S\b{\tau,0}\sb{\Lambda_\tau\sb{\hat{\rho}}}\\=\Upsilon \tau^3+\Xi\tau^4+O\b{\tau^5}~~,
\end{multline}
where the initial state is denoted by $\hat{\rho}\equiv\hat{\rho}_S\b 0\otimes\hat{\rho}_E\b 0$.
The first and second order terms vanish and the higher order terms are given by
\begin{gather}
    \Upsilon=\f{i}{\hbar^{3}}\text{tr}_{E}\b{\b{\hat H'+\hat H_{S}}\hat \rho\hat X-\hat X\hat \rho\b{\hat H'+\hat H_{S}}} \label{eq:higher_order_terms}\\
    \Xi=\f{1}{\hbar^{2}}\text{tr}_{E}\b{\hat X\hat \rho \hat H'\hat H_{S}-\hat H'\hat H_{S}\hat \rho\hat X}~~.
   \nonumber
\end{gather}

We keep track of terms up to fourth order in $\tau$ to be consistent with the linearization of the propagators, Eq. \eqref{eq:infit_propagators}. In the applied Markovian framework, the higher order terms, $\Upsilon$ and $\Xi$, must vanish when the maps commute. In turn, this imposes two strict conditions on the form of the general Hamiltonian $\hat{H}'$, Eq. \eqref{eq:higher_order_terms}. We find that if an arbitrary system state $\hat{\rho}_S\b t$ and a stationary environment state $\hat{\rho}_E$ are considered, except for pathological cases, the commutativity of the maps implies that $\hat{X}$ must vanish and strict energy conservation holds \footnote{ Note that for a general system state $\hat{\rho}_S\b 0$ the two terms in $\Upsilon$ may lead to different system operators. Meaning, that to satisfy the equality in Eq. \eqref{eq:higher_order_terms} both contributions must vanish.}. 

Overall, in the Markovian regime, commutativity of $\Lambda_t$ and ${\cal U}_S\b{t,0}$ suggests that the composite dynamics are generated by a strict energy conserving Hamiltonian. In addition it reflects
time translational symmetry \cite{marvian2014extending}.
Hence, there is a close connection between strict energy conservation and the commutativity property of the dynamical maps.

The property that the generators ${\cal L}$ and ${\cal H}$ commute has been identified in certain derivations of the quantum master equation. Alicki \cite{alicki1976detailed} based his derivation on detailed balance and that ${\cal L} \hat \rho_S^{eq} =0$ obtaind $[{\cal L},{\cal H}]=0$. 

Another derivation based on elastic repeated interactions or collision model leads to  both commutativity of the maps and strict energy conservation: $\sb{\hat S,\hat H_S+\hat H_E}=0$, 
where  $\hat S$ is the scattering matrix.
Each individual interaction represents a scattering event of the system with a single particle, with an associated map \cite{lippmann1950variational,newton2013scattering}
\begin{equation}
  \hat \rho_S(\tau)=\Lambda_{\tau}\sb{\hat{\rho}_S\b 0}= \text{tr}_E\b{\hat{S}\hat{\rho}_S\b 0\otimes \hat{\rho}_E \hat{S}^\dagger}~~.  
\end{equation}
Markovian dynamics are obtained in the low density limit \cite{karplus1948note}, when the time between collisions is much larger than the collision time $\tau \gg \tau_{s}$. This assumption softens the energy conservation condition to  a width of $\Delta E \sim h/\tau$.
The resulting master equation shares the same form as Eq. \eqref{eq:cal_D_final} \cite{dumcke1985low,Gisin2002}.
We emphasize that the derivation 
leading to the standard GKLS form does
not restrict the
collision strength \cite{de2018reconciliation}.

A GKLS master equation which satisfies the commutativity property has been obtained employing the weak 
system-bath coupling and the secular approximation. This procedure is commonly known as the Davies construction \cite{spohn1978irreversible,davies1974markovian}. In the setting of this derivation, the secular approximation acts as a filter which enforces energy conservation. This procedure is also equivalent to a coarse-graining in time, relaxing the energy conservation to $\Delta E \sim h/\tau_R$, where $\tau_R$ is the relaxation time \cite{cohen1998atom,elouard2020thermodynamics,mozgunov2020completely}.
In these constructions the connection between the commutativity property 
of the generators and the energy conservation suggests that such master equations, which satisfy the commutativity property, implicitly imply strict energy conservation.

\section{Discussion}
\label{sec:discussion}

Deriving the quantum master equation from first principles has been a subject of active research for more than 70 years.
The original derivations were based on either the collision model \cite{karplus1948note} or the weak coupling limit, introduced by Bloch and Redfield \cite{wangsness1953dynamical,redfield1957theory}. 
An major breakthrough in the field was the development of the GKLS generator \cite{lindblad1976generators,gorini1976completely}
based on postulates of a CPTP map \cite{kraus1971general}. The outcome was a rigid 
structure which enabled an easy route to construct models of physical reality. The caveat of this method is that these models could potentially be in conflict with thermodynamical laws.

This conflict initiated an effort to derive
a GKLS generator consistent with thermodynamics.
A pioneering study by Alicki imposed detailed balance conditions on the kinetic coefficients \cite{alicki1976detailed,alicki1977markov}.
Alternatively starting from the Born-Markov approximation and imposing the secular approximation, Davies constructed a thermodynamically compatible GKLS equation \cite{davies1974markovian}. In the Davies construction the generators of the unitary and dissipative part commute, and are therefore compatible with our present form \cite{spohn1978irreversible,roga2010davies}.

The present approach in comparison imposes additional
theromynamically inspired postulates on the dynamical CPTP map. These in turn lead to the restricted structure of the generator ${\cal L}$.  The advantage of the Davies construction is that for a specific model the kinetic coefficients $\gamma_k$ are determined completely, (not only up to a scale as in Eq. \eqref{eq:cal_D_final}).
The kinetic coefficients depend on the Bohr frequencies $\{\omega\}$ and the environment density of states at the corresponding energy. Specifically, the kinetic coefficients of the  pure-dephasing terms are related to a vanishing Bohr frequency. In the typical physical scenario, the density of states vanishes when $\omega\ra 0$, nulling the pure-dephasing in a second order expansion \cite{breuer2002theory}.
In a perturbation treatment in the  system-environment coupling, pure-dephasing appears only beyond fourth order in the perturbation expansion \cite{laird1991quantum}. In contrast, in the current analysis the pure-dephasing term emerges naturally from the unitary invariant set. 

A disadvantage of the Davies construction is that the conducted approximations limits the validity regime of the analysis. Specifically, 
the secular approximation is justified only when the course graining-time is much larger than the timescale related to the difference between the systems Bohr frequencies. However, in many physical systems,  such as in many-body systems, the level spacing can decrease to a point where the secular approximation breaks down. In these cases, it not clear how to proceed and obtain a GKLS master equation from first principles. Our construction hints that the oscillating phases may be an artifact of the approximations and constructions utilized in the microscopic derivation. In contrast to the microscopic derivation, the present construction does not restrict the magnitude of the Bohr frequencies and determines the unitary non-invariant Lindblad jump operators uniquely, as long as the Bohr frequencies remain non-degenerate.

The formal presented structure sheds light on unresolved issues in the discipline of open system dynamics:
\paragraph*{Local vs. global approaches} The microscopic derivations of Master equation can be categorized into local and global approaches. The local approach assumes the environment couples only locally to a part of the system. While in the global approach, the environment couples to global degrees of freedom of the whole system. The two approaches have been studied  and compared thoroughly \cite{levy2014local,barra2015thermodynamic,hofer2017markovian,gonzalez2017testing,de2018reconciliation,hewgill2020quantum,scali2020local,vadimov2020validity}.  In the presented construction the obtained generator coincides with the global master equation. Meaning that, the unitary  non-invariant operators $\{\hat{F}\}$ which constitute the jump operators of the master equation are eigenoperators of the ``global" system propagator and not solely of a sub-system. 

The origin of this property can be traced back to the strict energy conservation condition. For instance, consider a system which is decomposed of two parts, a part which interacts with the environment $\hat{H_S}^{\b{i}}$ and a passive part $\hat{H_S}^{\b{p}}$. When the two parts are coupled, the system Hamiltonian reads 
$\hat{H}_S=\hat{H_S}^{\b{i}}+\hat{H_S}^{\b{p}}+\hat{H_S}^{\b{c}}$. When $|\hat H_{S}^{\b{c}}| \ll |\hat{H_S}^{\b{i}}|,|\hat{H_S}^{\b{p}}|$ it would seem appropriate to describe the dynamics by a local Master equation,
where the generator acts only on a part of the system. However, this construction violates strict energy
conservation since the internal and external interactions terms do not commute. In contrast, a global master equation manifests 
strict energy conservation for the combined system.
It has been shown recently that thermodynamics consistency can be restored by adding a resource, a work term to compensate for interaction \cite{hewgill2020quantum}. For example an additional local dephasing term on the passive system would do the job.

\paragraph*{Existence of coherence in heat transport}
The present analysis can also be extended to transport phenomena, where the primary system is coupled to several
environments characterized by different temperature.
In this case, we can combine
different environments which are all together in a stationary state.
As a result, the GKLS generator \eqref{eq:gen_GKS} will  have kinetic coefficients which are
combinations of the detailed balance coefficients of individual environments. 
As a consequence, the system steady state lacks coherence in the energy basis. This observation has been noticed in the analysis of a three-qubit absorption refrigerator \cite{correa2013performance}. In contrast,
the local treatment exhibited coherence 
\cite{brunner2014entanglement}.

\paragraph*{Exceptional points in open system dynamics}
Exceptional points (or non-Hermitian degeneracies) \cite{heiss1990avoided,am2015exceptional,moiseyev1980association,moiseyev2011non} appear when two or more eigenvalues coalesce and the associated eigenvectors become linearly dependent at this point. 
Exceptional point can be associated with ${\cal{PT}}$-symmetry breaking \cite{heiss2012physics} and non-exponential decay, which was observed in optics and open-quantum systems \cite{am2015exceptional,naghiloo2019quantum,pick2019robust}.


At the exceptional point the matrix becomes defective and consequently cannot be diagonalized. Nevertheless, they can be cast by a similarity transformation into a Jordan normal form.  Jordan matrices do not commute with an Hermitian matrix which differs from the identity.
In Sec. \ref{sec:restricted_sturcture_of_ME} we found that the dynamical maps $\Lambda$ and ${\cal U}_S$ commute. This implies that $\Lambda$ and $\cal L$ are normal matrices and therefore do not have exceptional points.  Thus, Theorem 1 serves as a no-go theorem (under the considered postulates) for the existence of exceptional points in the dynamics a quantum system interacting with an initial stationary environment.

The present construction allows modifying the unitary part of the master equation. For example, one can add the Lamb-shift Hamiltonian which commutes with the dynamical map of the free propagation. In turn, this will lead to shifts in energy in the system Hamiltonian, which will slightly modify the kinetic coefficients without changing the Lindblad jump operators \cite{vadimov2020validity}. 

\paragraph*{Definition of heat current}  The postulate of strict energy conservation unambiguously  defines the heat current as $\dot{Q}=\mean{{\cal{L}}\sb{\hat{H}_S}}$. Since the change of energy in the system corresponds to energy transferred to the environment. This heat transport accounts for entropy production in the environment $\Delta S_E$ due to energy transfer. Additional entropy production is generated by loss of coherence and mutual information \cite{lindblad1973entropy,goold2016role}.

Overall, the theme of this paper is the formulation of thermodynamic principles in terms of mathematical statements imposed on the quantum dynamical map. Through spectral analysis we have obtained a restricted structure of the master equation which by construction complies with thermodynamics. This structure can be employed as an independent validator of open system dynamics which are obtained by means of physical approximations.


\begin{acknowledgements}
 We thank Robert Alicki, Peter Salamon, James Nulton
Erez Boukobza, Nimrod Moiseyev , Adi Pick, Ander Tobalina, Amikam Levy, Daniel Lidar and Luis Correa
for sharing their views and comments.
This research was supported by the Adams Fellowship  Program of the Israel Academy of Sciences and Humanities, the National Science Foundation under Grant No. NSF PHY-1748958 and The Israel Science Foundation Grant No.  2244/14.
\end{acknowledgements}

\appendix

\section{Mathematical construction of the quantum dynamical semi-group generator and the orthogonality of the Lindblad operators}
\label{apsec:construction_from_map}
The present section includes a summarized textbook construction of the general form of the generator $\cal L$, for further details Cf. Ref. \cite{breuer2002theory} chapter 3.2 which is in the spirit of the original work of Gorini Kossakowski and Sudarshan \cite{gorini1976completely}. We utilize a number of the intermediate results of this proof in the main derivation. 

We begin by introducing an orthonormal operator basis $\{\hat{S}\}$ for the system's Liouville space. The operators of the basis are chosen such that, a single operator is proportional to the identity, $\hat{S}_{N^2}=\b{1/ N}^{1/2}\hat{I}_S$ and the others are traceless.  
Utilizing the completeness relation and the inner product in Liouville space, the Kraus operators Eq. \eqref{eq:Kraus_form} can be expressed as 
\begin{equation}
    \hat{K}_{\mu \nu}=\sum_{k=1}^{N^2}\b{\hat{S}_i,\hat{K}_{\mu \nu}}\hat{S}_i~~.
    \label{eq:ap1}
\end{equation}
Substituting \eqref{eq:ap1} into the dynamical map Eq. \eqref{eq:Kraus_form} we obtain
\begin{equation}
\Lambda_t\sb{\hat{\rho}_S\b 0}=\sum_{i,j=1}^{N^2}r_{ij}\hat{S}_i\hat{\rho}_S\b 0\hat{S}_j^\dagger ~~,
\label{eq:ap2}
\end{equation}
 where 
\begin{equation}
    r_{ij} = \sum_{\mu,\nu}\b{\hat{S}_{i},\hat{K}_{\mu \nu}}\b{\hat{S}_j,\hat{K}_{\mu \nu}}^*
    \label{eqap:r_ij}
\end{equation}
The coefficient matrix $r=\sb{r_{ij}}$ can be shown to be Hermitian and positive.

Next, we introduce the following coefficients
\begin{gather}
 a_{N^2N^2}=\lim_{\eps\ra 0}\f{r_{N^2N^2\b{\eps}}-N}{\eps}   \label{eqap:a_coeff}\\  
  a_{iN^2}=\lim_{\eps\ra 0}\f{r_{iN^2}\b{\eps}}{\eps}\\  
  \nonumber
  a_{ij}=\lim_{\eps\ra 0}\f{r_{ij}\b{\eps}}{\eps}~~,
  \nonumber
\end{gather}
where $i,j=1,\dots,N^2-1$, and the operators $\hat{S}=\b{1/N}^{1/2}\sum_{i=1}^{N^2-1}a_{iN^2}\hat{S}_i$, $\hat{R}=\f{1}{2N}a_{N^2N^2}\hat{I}_S+\f{1}{2}\b{\hat{F}^\dagger +\hat{F}}$ and $\hat{J}=\f{1}{2i}\b{\hat{F}^\dagger-\hat{F}}$, with $\hat{F}=\f{1}{\sqrt{N}}\sum_{i=1}^{N^2-1}a_{iN^2}\hat{F_i}$.

Substituting the coefficients of Eq. \eqref{eqap:a_coeff} into the generator of the dynamical semi-group  ${\cal L}\sb{\hat{\rho}_S} =\lim_{\eps\ra0}\f{1}{\eps}\b{{\Lambda}_\eps\sb{\hat{\rho}_S}-\hat{\rho}_S}$  and expressing the result in terms of $\hat{S}$, $\hat{R}$ and the Hermitian operator $\hat{J}$, the generator can be written as
\begin{multline}
    {\cal L}\sb{\hat{\rho}_S\b t}=-i\sb{\hat{J},\hat{\rho}_S\b t}+\left\{\hat{R},\hat{\rho}_S\b t \right\} 
    \\+\sum_{ij=1}^{N^2-1}a_{ij}\hat{S}_i\hat{\rho}_S\b t\hat{S}_j^\dagger~~.
    \label{eqap:L_before}
\end{multline}
This form must preserve the trace of the density matrix $\text{tr}_S\b{{\cal{L}}\sb{\hat{\rho}_S\b t}}=0$, which implies that $\hat{R}=-\f{1}{2}\sum_{i,j}^{N^2-1}a_{ij}\hat{S}_j^\dagger\hat{S}_i$. Inserting this relation into Eq. \eqref{eqap:L_before} leads to the form originally obtained by  Gorini, Kossakowski and Sudarshan \cite{gorini1976completely}
\begin{multline}
    {\cal L}\sb{\hat{\rho}_S\b t}=-i\sb{\hat{J},\hat{\rho}_S\b t}\\
    +\sum_{i,j=1}^{N^2-1}a_{ij}
    \b{\hat{S}_i\hat{\rho}_S\b t\hat{S}_j^\dagger-\f{1}{2} \left\{\hat{S}_j^\dagger\hat{S}_i ,\hat{\rho}_S\b t\right\}}~~.
\end{multline}

The GKLS form is obtained by basis transformation which diagonalizes the coefficient matrix $a$.
The relations defined in Eq. \eqref{eqap:a_coeff} and the properties of the  coefficient matrix $c$ imply that  $a=\sb{a_{ij}}$ is also Hermitian and positive. As a consequence, it can be diagonalized by a similarity transformation $u^\dagger a u=\text{diag}\b{\lam_1,\dots,\lam_{N^2-1}}$. Introducing a new set of operators $\{\hat{L}\}$, satisfying $\hat{S}_i=\sum_{k}^{N^2-1}u_{ik}^*\hat{L}_k$ and expressing $\cal L$ in terms of these operators leads to the final general GKLS form 
\begin{multline}
     {\cal L}\sb{\hat{\rho}_S\b t}=-i\sb{\hat{J},\hat{\rho}_S\b t}\\
    +\sum_k \lambda_k\b{ \hat{L}_k\bullet \hat{L}_k^\dagger-\f{1}{2}\left\{\hat{L}_k^\dagger \hat{L}_k,\bullet \right\}}~~.
\end{multline}

\section{Restriction on the Master equation}
\label{apsec:restrictions_on_the_master_eq}
 In section \ref{sec:restricted_sturcture_of_ME} we  express the general dissipative term of the generator $\cal{D}^*$ Eq. \eqref{eq:gen_GKS} in terms with jump operators from $\mathfrak{K}=\{\hat{F}\}\bigcup\{\hat{\Pi}\}$. 
 In this form, the terms of the sum are generally of the form 
 \begin{multline}
      {\cal D}_{nmn'm'}^*\sb{\bullet}\equiv\hat{S}_j^\dagger\bullet\hat{S_i}-\f{1}{2}\{\hat{S}_j^\dagger\hat{S}_i,\bullet \}\\
      =\ket n\bra m\bullet\ket{n'}\bra{m'}-\f 12\{\ket n\braket m{n'}\bra m',\bullet\} 
      \label{eq:D_nmnm_term}
 \end{multline}
 for $n,m,n',m'\in{1,\dots,N}$. 
 For a general operator $\ket{k}\bra{l}$ in $\mathfrak{K}$ Eq. \eqref{eq:D_nmnm_term} becomes
\begin{multline}
     {\cal D}_{nmn'm'}^*\sb{\ket{k}\bra{l}}=\delta_{mk}\delta_{ln'}\ket{n}\bra{m'}\\
     -\delta_{mn'}\f{1}{2}\b{\delta_{m'k}\ket{n}\bra{l}+\delta_{nl}\ket{k}\bra{m'}}~~.
     \label{eq:D_nmnm_term_specific}
\end{multline}

 Assuming the spectrum of $\cal{U}_S$ is non-degenerate, the commutativity of $\cal D$ and ${\cal H}_S$, Eq. \eqref{eq:commutivity}, implies the eigenoperator condition 
 \begin{equation}
 {\cal{D}}^*\sb{\hat{F}_k}\propto \hat{F}_k~~.  
 \label{eqap:eignenoperator_cond}
 \end{equation}
 This induces strict restrictions on the components ${\cal D}_{nmn'm'}^*$ which can appear in $\cal D^*$. To analyse which component contribute to $\cal D^*$ it is convenient to represent the dissipator in Liouville space. The elements of the superoperator in Liouville space in the basis $\mathfrak{F}$ are defined as $d_{ij}=\text{tr}\b{\hat{S}_i^\dagger {\cal D}\sb{\hat{S}_j}}$. Since, $\{\hat{F}\}$ are eigenoperators of $\cal D^*$ in , $\widetilde{\cal D}$ obtains the form
 \begin{equation}
    \widetilde{\cal D}^*=\sb{\begin{array}{cc}
\tilde{\Delta}_{v} & 0\\
0 & \tilde{\Delta}_{i}
\end{array}}~~,
\label{eq:tilde_D}
\end{equation}
where $\tilde{\Delta}_{v}$ is diagonal and $\tilde{\Delta}_{i}$  is a diagonalizable matrix. 
The form of $\widetilde{\cal D}^*$ highlights that the unitary invariant and non-invariant components are independent
\begin{equation}
    {\cal{D}}^*\sb{\hat{\Pi}_n}=\sum_{nm}b_{nm}\hat{\Pi}_m~~.  
    \label{eqap:cond_2}
\end{equation}

Relations \eqref{eqap:eignenoperator_cond} and \eqref{eqap:cond_2} greatly restrict the term ${\cal D}_{nmn'm'}^*$ included in the dissipator. We analyse the different cases:
For $n\neq m'$ only the first term in Eq. \eqref{eq:D_nmnm_term} contributes, condition \eqref{eqap:eignenoperator_cond} then implies that $n=k$ and $m'=l$. Hence, for $n\neq m'$ only terms of the form ${\cal D}_{kkll}^*$ contribute to $\cal D^*$. One can maybe suspect that somehow a combination of terms can cancel eachother allowing for condition \eqref{eqap:eignenoperator_cond} to be satisfied without nulling ${\cal D}_{nmn'm'}^*$ components with do not comply with Eq. \eqref{eqap:eignenoperator_cond} and Eq. \eqref{eqap:cond_2}. However, this cannot be, since each component is associated with a different non-diagonal matrix element  $d_{ij}\equiv d_{klk'l'}=\text{tr}\b{\ket{l}\bra{k} {\cal D}\sb{\ket{k'}\bra{l'}}}$ with $k\neq k'$ and $l\neq l'$.
When $m=n'$ and $k=l$ (substituting the projection operator $\hat{\Pi}_k$ into ${\cal{D}}_{nmmm'}$) Eq.   \eqref{eqap:cond_2} suggests that we must have $m=n'$. Thus, only terms of the form ${\cal{D}}_{kllk}^*$.

Overall, the restrictions imply that only two possible components can contribute to the sum ${\cal D}_{kkll}^*$ and ${\cal D}_{kllk}^*$, leading to restricted form of Eq. \eqref{eq:cal_D_pre}.

\section{Properties of the matrix $\alpha$}
\label{apsec:properties_of_alpha}

To study the properties of the matrix $\alpha$ we first analyze certain elements of the $r$ matrix. This is achieved by building upon the general construction presented in Appendix \ref{apsec:construction_from_map}

 We choose the orthonormal basis $\{\hat{S}\}$ to be composed of the transition operator of $\{\hat{F}\}$ and an orthonormal set $\{\hat G\}$ which spans the unitary  invariant space of ${\cal U}_S$. Specifically, we choose the basis of the unitary  invariant subspace $\{\hat{G}\}$ to be proportionate to the diagonal Gell-Mann matrices \cite{bertlmann2008bloch}. As required the Gell-Mann matrices are traceless, but may have to be re-normalized. We therefore scale them so the basis $\mathfrak{F}$ has the same properties as $\{\hat{S}\}$ in  Appendix \ref{apsec:construction_from_map}. 
The identification of $\{\hat{G}\}$ with the Gell-Mann matrices (up to a constant) allows expressing the unitary  invariant basis operator in terms of the energy projection operators
\begin{equation}
  \hat{G}_k=\sum_l g_{kl}\hat{\Pi}_l~~,  
  \label{eqap:G_k_decomp}
\end{equation}
where the weights $g_{kl}$ are real.

We utilize this relation to analyze the properties of coefficients $r_{ij}$.
Inserting, Eq. \eqref{eqap:G_k_decomp} into the coefficients $r_{ij}$, Eq. \eqref{eqap:r_ij}, associated with $\hat{S}_i,\hat{S}_j\in\{\hat{G}\}$ leads to
\begin{multline}
    r_{G_i G_j} = \sum_{\mu,\nu}\b{\hat{G}_{i},\hat{K}_{\mu \nu}}\b{\hat{G}_j,\hat{K}_{\mu \nu}}^*\\=
 \sum_{l,l'}g_{il}g_{jl'}\sum_{\mu,\nu}\b{\hat{\Pi}_{l},\hat{K}_{\mu \nu}}\b{\hat{\Pi}_{l'},\hat{K}_{\mu \nu}}^*~~.
\end{multline}

Utilizing the inner-product in Liouville space the sum over inner products can be expressed as 
\begin{multline}
   \varXi\equiv\sum_{\mu,\nu} \b{\hat{\Pi}_{l},\hat{K}_{\mu \nu}}\b{\hat{\Pi}_{l'},\hat{K}_{\mu \nu}}^*\\=\sum_{\mu\nu}\bra l\hat{K}_{\mu\nu}\ket l\bra{l'} \hat{K}_{\mu\nu}^{\dagger}\ket{l'}~.
\end{multline}
Substituting the explicit form of the Kraus operators  $\hat{K}_{\mu\nu}=\sqrt{\lam_{\mu}}\bra{\chi_{\nu}}\hat U\b{t,0}\ket{\chi_{\mu}}$ (Eq. \eqref{eq:Kraus_form}) lead to 
\begin{multline}
    \varXi = \sum_{\mu\nu}\bra{i,\chi_{\nu}}\hat{U}\b{t,0}\ket{\chi_{\mu},i}\bra{j\chi_{\mu}}\hat{U}\b{t,0}\ket{\chi_{\nu},j}\\= \sum_{\nu}\bra{i,\chi_{\nu}}\hat{U}\b{t,0}\ket{i}\b{\sum_\mu{\ket{\chi_{\mu}}}\bra{\chi_{\mu}}}\bra{j}\hat{U}\b{t,0}\ket{\chi_{\nu},j}\\=\text{tr}_E\b{\bra{i}\hat{U}\b{t,0}\ket{i}\bra{j}\hat{U}\b{t,0}\ket{j}}=\varXi ^*.
    \label{eqap:varXi}
\end{multline}
The third equality is implied by the completeness relation and the definition of the partial trace. The fourth equality stems from the unitarity of the propagator.
Equation \eqref{eqap:varXi} implies that $\varXi $ is real.

We can now conclude that the coefficient matrix of the unitary  invariant sub-space $r^{inv}=\sb{r_{G_i G_j}}$ is a real Hermitian matrix, or equivalently, a symmetric matrix. As a result, the associated coefficient matrix $a^{inv}=\sb{a_{G_i G_j}}$ (Eq. \eqref{eqap:a_coeff}) is also symmetric. 
In section \ref{sec:restricted_sturcture_of_ME}, the coefficient matrix $\alpha$ is obtained by a basis transformation, form $\{\hat{G}\}$ to $\{\hat{\Pi}\}$, this implies that $\alpha$ can be diagonalized by a similarity transformation by means of an orthogonal matrix $Q$: $Q^T \alpha Q=\text{diag}\b{\lam_1,\dots,\lam_N}$.
The property that the elements of $Q$ are real suggest to any operator of the form $\hat{V}_n=\sum_{n}Q_{ni}\hat{\Pi}_n$ is Hermitian.

\section{Detailed balance by means of the generator's fixed point}
\label{apsec:kinetic_coeff}
The fixed point of the $\cal L$ determines a detailed balance relation between two dependent coefficients $\gamma_k$ and $\gamma_{-k}$, Eq. \eqref{eq:cal_D_final}.
To show this property we calculate the action of the generator on the fixed point and enforce the condition for which  ${\cal L}\sb{\hat{\rho}_S^{th}}=0$. 
The fixed point $\rho_S^{th}=Z_S^{-1}e^{-\beta \hat{H}_S}$ trivially commutes with $\hat{H}_S$ which simplifies the invariance condition to  ${\cal D}\sb{\hat{\rho}_S^{th}}=0$.

The dissipative term is composed from two sums. The first sum includes the unitary non-invariant operators $\{\hat{F}\}$ of ${\cal U}_S$ and induces transitions between the energy levels.  The second sum is composed of unitary invariant operators and generates pure-dephasing. For simplicity we introduce a short notation for these sums ${\cal D}={\cal D}_n+{\cal D}_i$.
The pure-dephasing component of the dissipator ${\cal D}_i$ is composed of energy projection operators $\{ \hat{\Pi}\}$. With the help of the relation 
\begin{equation}
    \hat{\Pi}_{k}^{\dagger}\hat{\Pi}_{r}\hat{\Pi}_{l}-\f 12\{\hat{\Pi}_{k}^{\dagger}\hat{\Pi}_{l},\hat{\Pi}_{r}\}=0
    \label{eqap:Pi}
\end{equation}
we can infer that ${\cal D}_i\sb{\hat{\rho}_S^{th}}=0$. In turn, the invariance condition then becomes
\begin{equation}
    {\cal D}_v\sb{e^{-\beta \hat{H}_S}}=0~~.
    \label{eqap:D_n}
\end{equation}

The Lindblad jump operators of ${\cal D}_n$ are transition operators between the energy level of $\hat{H}_S$. This property suggests that 
\begin{equation}
    \sb{\hat{H}_S,\hat{F}_{nm}}=-\hbar \omega_{nm}~~,
    \label{eqap:com1}
\end{equation}
where $\omega_{nm}=\b{\eps_m-\eps_n}/\hbar$ is the Bohr frequency between energy levels $\ket{n}$ and $\ket{m}$.
Utilizing the commutation relations and the Baker-Campbell-Hausdorff formula we deduce that
\begin{equation}
    \hat{F}_{nm}e^{-\beta \hat{H}_S} =e^{-\beta \hat{H}_S}\hat{F}_{nm}e^{-\beta \hbar \omega_{nm}}~~.
    \label{eqap:relation}
\end{equation}
We now utilize this relation to calculate ${\cal D}\sb{e^{-\beta \hat{H}_S}}$ assuming general kinetic coefficients.

The $k$'th term of ${\cal{D}}_v$, is constructed from two terms ${\cal D}^{k}_v={\cal B}_v^{k}+{\cal B}_v^{-k}$ (recall that $k$ runs over double indices $n$ and $m$) with
\begin{equation}
   {\cal B}_v^{k}\sb{e^{-\beta \hat{H}_S}} = \gamma_k\b{\hat{F}_{nm}e^{-\beta \hat{H}_{S}}\hat{F}_{nm}^{\dagger}-\f 12\left\{\hat{F}_{nm}^{\dagger}\hat{F}_{nm},e^{-\beta \hat{H}_{S}}\right\}}
\end{equation}
and
\begin{equation}
   {\cal B}_v^{-k}\sb{e^{-\beta \hat{H}_S}}  = \gamma_{-k}\b{\hat{F}_{nm}^{\dagger}e^{-\beta \hat{H}_{S}}\hat{F}_{nm}-\f 12\{\hat{F}_{nm}\hat{F}_{nm}^{\dagger},e^{-\beta \hat{H}_{S}}\}}
\end{equation}
With the help of relation \eqref{eqap:relation}  and its Hermitian conjugate we find that
\begin{gather}
    {\cal B}_v^{k}\sb{e^{-\beta \hat{H}_S}}=\gamma_k\b{e^{-\beta\omega_{nm}}\hat F_{nm}\hat F_{nm}^{\dagger}-\hat F_{nm}^{\dagger}\hat F_{nm}}e^{-\beta \hat{H}_{S}}
    \label{eqap:eqE7}\\
    {\cal B}_v^{-k}\sb{e^{-\beta \hat{H}_S}}=\gamma_{-k}\b{e^{\beta\omega_{nm}}\hat F_{nm}^\dagger \hat F_{nm}-\hat F_{nm}\hat F_{nm}^{\dagger}}e^{-\beta \hat{H}_{S}}
    \nonumber
\end{gather}
Since, terms ${\cal D}^{k}_v$ and ${\cal D}^{k'}_v$ are independent, condition \eqref{eqap:D_n} translates to $N\b{N-1}/2$ independent requirements 
\begin{equation}
   {\cal D}_v^{k}\sb{e^{-\beta \hat{H}_S}}={\cal B}_v^{k}\sb{e^{-\beta \hat{H}_S}}+{\cal B}_v^{-k}\sb{e^{-\beta \hat{H}_S}}=0~~.
   \label{eqap:indep_cond}
\end{equation}
Inserting Eq. \eqref{eqap:eqE7} into condition \eqref{eqap:indep_cond} leads to the detailed balance relation between the kinetic coefficients
\begin{equation}
    \gamma_k = \gamma_{-k}e^{\beta \omega_{nm}}~~.
\end{equation}

%

\end{document}